\documentstyle[aps,preprint]{revtex}
\begin{document}
\draft
\title{Baryogenesis in Fresh inflation}
\author{Mauricio Bellini\footnote{E-mail address: mbellini@mdp.edu.ar;
bellini@ginette.ifm.umich.mx}}
\address{Instituto de F\'{\i}sica y Matem\'aticas, \\
Universidad Michoacana de San Nicol\'as de Hidalgo, \\
AP:2-82, (58041) Morelia, Michoac\'an, M\'exico.}

\maketitle
\begin{abstract}
I study the possibility of baryogenesis can take place in fresh inflation.
I find that it is possible that violation of baryon number conservation
can occur during the period out-of-equilibrium in this scenario. Indeed,
baryogenesis could be possible
in the range of times ($10^{9} - 10^{12}$) ${\rm G^{1/2}}$, before
the thermal equilibrium is restored at the end of fresh inflation.
\end{abstract}
\vskip 2cm
\noindent
{\rm Pacs:} 98.80.Cq \\
\vskip 2cm
{\em Introduction: }
The idea of inflation is one of the most reliable concepts in modern
cosmology\cite{1,11,2,bcms}.
It can solve the horizon and flatness problem in
standard big bang cosmology and also provide us with the seeds of the
large scale structure. The standard inflationary period
proceeds while a scalar field called an inflaton slowly evolves
along a sufficiently flat potential.
The standard slow - roll inflation model separates expansion and reheating
into two distinguished time periods. It is first assumed that
exponential expansion from inflation places the universe in a supercooled
second order phase transition. Subsequently thereafter the universe is
reheated. Two outcomes arise from such a scenario. First,
the required density perturbations in this cold universe are left to be
created by the quantum fluctuations of the inflaton. Second, the
scalar field oscillates near the
minimum of its
effective potential and produces elementary particles.
Inflation has improved upon the scenario for generating a baryon excess
in two ways. First, it has imposed the condition that the initial baryon
asymmetry vanishes as a prediction of just as a convenient assumption.
Second, has provided a maximum temperature below which the baryon generation
process must take place. For standard inflation
baryon number violation must take
place at temperatures below the reheating temperature after inflation
if our present universe is to end up with a non-zero baryon density.
The picture of baryon generation which inflation give us is the following.
When inflation ends, a scalar field which had been
storing the energy density which drove
inflation in the form of a cosmological
constant is suddenly released from a meta-stable configuration and it
rapidly proceeds to the minimum of its potential energy. As it approaches the
minimum the potential energy of the meta-stable state is transferred
to kinetic energy and the field value oscillates around the minimum of the
potential. These field oscillations are completely equivalent to a coherent
state of the particles corresponding to the scalar field. The next
step is that these particles decay, or equivalently, that field oscillations
are damped by the production of other particles which are coupled to
the oscillating field. This accomplishes the reheating of the universe\cite{21}.
Baryon number is generated when the temperature of the universe
dips below the mass of some suitable particle. There is a
temporary loss of equilibrium when that particle cannot decay quickly
enough to reduce its density in response to the rapidly cooling universe.
The particle being discussed must have decay modes which violate both
CP and baryon number conservation. There are very
good reasons to suspect that GUT baryogenesis does not occur if this is
the way reheating happens. The main reason is that density and temperature
fluctuations observed in the present universe require the inflaton
potential to be extremely flat. This
means that the couplings of the inflaton field
to the other degrees of freedom cannot be too large, since large couplings
would induce large loop corrections to the inflaton potential,
spoiling its flatness. As a result, the radiation temperature is expected
to be smaller than $10^{14} \  {\rm GeV}$ by several orders of magnitude\cite{rio}.

An interesting idea called preheating was introduced more recently\cite{KLS}.
When the inflaton field oscillates around the minimum of the potential
the Klein-Gordon equation for the modes can be cast in the form of a
Mathieu equation. A crucial observation for baryogenesis is that particles
with mass larger than that of the inflaton may be produced during preheating\cite{KLR}.

Recently a new model of inflation called {\em fresh inflation}
was proposed\cite{22}. 
This one has the following characteristics:

(a) The universe begins from an unstable primordial matter field
perturbation with energy density nearly $M^4_p$ and chaotic
initial conditions. Initially the universe there is no thermalized
[$\rho_r(t=t_0)=0$]. Later, the universe describes a second order
phase transition, and the inflaton rolls down towards its minimum
energetic configuration.
(b) Particles production and heating occur together during the rapid
expansion of the universe, so that the radiation energy density
grows during fresh inflation ($\dot\rho_r >0$). The Yukawa interaction
between the inflaton field and other fields in a thermal bath
lead to dissipation which is responsible for the slow rolling of the
inflaton field. So, the slow-roll conditions are physically justified
and there are not a requirement of a nearly flat potential in fresh
inflation.
(c) There is no oscillation of the inflaton field around the minimum
of the effective
potential due to the strong dissipation produced by the Yukawa
interaction ($\Gamma \gg H$). This fact also provides thermal equilibrium
in the last phase of fresh inflation.

In this work I study baryogenesis in fresh inflation. During the
early period of fresh inflation (when $\Gamma \ll H$) there is not
thermal equilibrium, but after $t>t_E$ the thermal
equilibrium es restored (once the condition $\Gamma \gg H$ is
fullfiled). The main subject of this paper is to study
the possibility that baryon asymmetry can take place
during the period before thermal equilibrium
is restored in fresh inflation. 

{\em Review of Fresh Inflation:}
I consider a Lagrangian for a $\phi$-scalar field minimally coupled
to gravity, which also interacts with another $\psi$-scalar field
by means of a Yukawa interaction,
\begin{equation}\label{1}
{\cal L} = - \sqrt{-g} \left[\frac{R}{16\pi G} +\frac{1}{2}
g^{\mu\nu} \phi_{,\mu}\phi_{,\nu} + V(\phi) +
{\cal L}_{int}\right],
\end{equation}
where $g^{\mu\nu}$ is the metric
tensor, $g$ is its determinant and $R$ is the scalar curvature. 
The interaction Lagrangian is given by 
${\cal L}_{int} \sim -{\rm g}^2 \phi^2 \psi^2$,
where $\psi$ is a scalar field in the thermal bath.
Furthermore,
the indices $\mu,\nu$ take the values $0,..,3$
and the gravitational constant is $G=M^{-2}_p$ (where
$M_p = 1.2 \times 10^{19} \  GeV$ is the Planckian mass).
The Einstein equations for a globally flat, isotropic, and homogeneous
universe described by a Friedmann-Robertson-Walker metric
$ds^2 = -dt^2 + a^2(t) dr^2$ are given by
\begin{eqnarray}
3 H^2 & =& 8\pi G\left[ \frac{\dot\phi^2}{2} + V(\phi) +
\rho_r\right], \label{4} \\
3H^2 + 2 \dot H & = & -8\pi G\left[\frac{\dot\phi^2}{2}-
V(\phi) + \rho_r\right], \label{5}
\end{eqnarray}
where $H={\dot a\over a}$ is the Hubble parameter and $a$ is the
scale factor of the universe. 
The overdot denotes
the derivative with respect to the time.
On the other hand, if $\delta=\dot\rho_r+4H\rho_r$
describes the interaction between the inflaton and the bath,
the equations of motion for $\phi$ and $\rho_r$
are
\begin{eqnarray}
&& \ddot\phi+3H\dot\phi+V'(\phi) + \frac{\delta}{\dot\phi}=0,\label{6}\\
&& \dot\rho_r+4H\rho_r-\delta=0. \label{7}
\end{eqnarray}
As in a previous paper\cite{22}, I will consider a Yukawa interaction
$\delta = \Gamma(\theta) \  \dot\phi^2$, where $\Gamma(\theta)=
{g^4_{eff}\over 192\pi}\theta$\cite{41}
and $\theta \sim \rho^{1/4}_r$
is the temperature of the bath.
If $p_t={\dot\phi^2 \over 2}+{\rho_r\over 3} - V(\phi)$
is the total pressure and $\rho_t=\rho_r+{\dot\phi^2 \over 2}+V(\phi)$ is the
total energy density, the parameter $F={p_t+\rho_t \over \rho_t}$
which describes the evolution of the universe during inflation\cite{81} is
\begin{equation}\label{8}
F= - \frac{2\dot H}{3 H^2} = \frac{\dot\phi^2+\frac{4}{3} \rho_r}{
\rho_r+ \frac{\dot\phi^2}{2}+V}>0.
\end{equation}
When fresh inflation starts (at $t=G^{1/2}$), the radiation energy 
density is zero, so that $F\ll 1$. 

I will consider the parameter $F$ as a constant.
From the
two equalities in eq. (\ref{8}), one obtains the following equations:
\begin{eqnarray}
&& \dot\phi^2 \left(1-\frac{F}{2}\right) +
\rho_r\left(\frac{4}{3}-F\right)-F \  V(\phi)=0, \label{9} \\
&& H= \frac{2}{3 \int F \  dt}.\label{10}
\end{eqnarray}
Furthermore, because of $\dot H = H'\dot\phi$ (here the prime denotes
the derivative with respect to the field), from the first equality
in eq. (\ref{8}) we obtain the equation that
describes the evolution for $\phi$,
\begin{equation}\label{11}
\dot\phi=-\frac{3 H^2}{2 H'}F,
\end{equation}
and replacing eq. (\ref{11}) in eq. (\ref{9}), the radiation energy
density can be described as functions of $V$, $H$ and $F$\cite{22}
\begin{equation}\label{12}
\rho_r = \left(\frac{3F}{4-3F}\right) V - \frac{27}{8}
\left(\frac{H^2}{H'}\right)^2 \frac{F^2(2-F)}{(4-3F)}.
\end{equation}
Finally, replacing eqs. (\ref{11}) and (\ref{12}) in eq. (\ref{4}),
the potential can be written as a function of the Hubble parameter
and $F$ (which is a constant)
\begin{equation}\label{13}
V(\phi) = \frac{3}{8\pi G} \left[\left(\frac{4-3F}{4}\right)
H^2 + \frac{3\pi G}{2} F^2\left(\frac{H^2}{H'}
\right)^2\right].
\end{equation}

Fresh inflation was proposed for a global group $O(n)$ involving
a single $n$-vector multiplet of scalar fields $\phi_i$\cite{wei},
such that making $(\phi_i\phi_i)^{1/2}\equiv \phi$, the effective
potential $V_{eff}(\phi,\theta)=V(\phi)+\rho_r(\phi,\theta)$ can be
written as
\begin{equation}\label{14}
V_{eff}(\phi,\theta) = \frac{{\cal M}^2(\theta)}{2} \phi^2+
\frac{\lambda^2}{4}\phi^4,
\end{equation}
where ${\cal M}^2(\theta) = {\cal M}^2(0)+{(n+2) \over 12} \lambda^2\theta^2$
and $V(\phi)= {{\cal M}^2(0)\over 2} \phi^2+{\lambda^2 \over 4}\phi^4$.
Furthermore, ${\cal M}^2(0) >0$ is the squared mass at
zero temperature, which is given by ${\cal M}^2_0$ plus renormalization
counterterms in the potential ${1 \over 2} {\cal M}^2_0 (\phi_i\phi_i)+
{1 \over 4} \lambda^2 (\phi_i\phi_i)^2$\cite{wei}.
I will take 
into account the case without symmetry breaking, ${\cal M}^2(\theta) >0$
for any temperature $\theta$, so that
the excitation spectrum consists of $n$ bosons with
mass ${\cal M}(\theta)$.
The effective potential (\ref{14}) 
is invariant under $\phi \rightarrow -\phi$ 
reflections and $n$ is the number of created particles due to
the interaction of $\phi$ with the particles in the thermal bath, such
that\cite{22}
\begin{equation}\label{15}
(n+2) = \frac{2\pi^2}{5\lambda^2}g_{eff} \frac{\theta^2}{\phi^2},
\end{equation}
because the radiation energy density is given by $\rho_r={\pi^2 \over 30},
g_{eff} \theta^4$, where $g_{eff}$ denotes the effective degrees
of freedom of the particles and it is assumed that $\psi$ has no                                                
self-interaction. A particular solution of eq. (\ref{13}) is
\begin{equation}
H(\phi) =4 \sqrt{\frac{\pi G}{3(4-3F)}} \  {\cal M}(0)  \  \phi,\label{16} \\
\end{equation}
where the consistence relationship implies:
$\lambda^2 = {12\pi G F^2 \over (4-3 F)} {\cal M}^2(0)$\cite{22}.
From eq. (\ref{10}), and due to 
$H=\dot a/a$, one obtains the scale factor
as a function of time
\begin{equation}\label{18}
a(t) \sim  t^{\frac{2}{3F}}.
\end{equation}
The number of $e$ folds $N=\int^{t_e}_{t_s} H(t) dt$ ($t_s$ and $t_e$
are the time when inflation start and ends) is given by
$N = \left.{2 \over 3F} {\rm ln}(t)\right|^{t_e}_{t_s}$.
With Planckian unities ($G^{-1/2} \equiv M_p=1$) inflation starts when
$t_s = G^{1/2} =1$. Hence, for $t_e \simeq 10^{13} \  G^{1/2}$, one
obtains $N>60$ for $F<1/3$. So, the condition $F< 1/3$ assures the
slow-rolling of the inflaton field during fresh inflation.
So, fresh inflation solve the problem of warm inflationary scenarios
considered in\cite{YL}.
Taking $g_{eff} \simeq 10^2$, ${\cal M}^2(0) = 10^{-12} \  {\rm M^2_p}$, and
$t_e \simeq 10^{13} G^{1/2}$ one obtains the number of
created particles at the end of fresh inflation
$n_e \simeq 10^{13}$. Furthermore, the time evolution of the
inflaton is given by $\phi(t) = \lambda^{-1} t^{-1}$\cite{22}.

{\em Baryogenesis in Fresh inflation:}
If the reheat temperature is sufficiently high, then baryogenesis
can proceed as it does in the standard cosmology, through the
out-of-equilibrium decays of superheavy bosons whose interactions
violate $B$, $C$ and $CP$ conservation\cite{7}.
As usual, $\epsilon$ is related to the branching ration of $\phi$
into channels which have net baryon number, and the $C$, $CP$ violation
in the $B$-nonconserving decay modes. For simplicity suppose that
only two decay channels have net baryon number equal to $B_1$ and $B_2$, then
$\epsilon \simeq (B_1 - B_2)(r_1-\bar r_1)(r_1+\bar r_2)$, where
$r_i$ ($\bar r_i$) is the branching ration into channel $i$ ($\bar i$).
The $C$, $CP$-violating effects involve higher - order loop corrections
$(r_1 - \bar r_1) \le O(\alpha) \ge 10^{-2}$, where $\alpha$ is the
coupling strength of the particle exchanged in the loop.

For the values of the parameters here adopted
the thermal equilibrium holds for $t > t_E\simeq 10^{12} \  {\rm G^{1/2}}$, so that
baryogenesis must take place before it. If baryon number is
not conserved there is no reason for the proton to be
stable and in fact, most theories which can produce a baryon
asymmetry also predict a finite lifetime for the proton. Experiments
now constrain this lifetime to be longer than $10^{32}$ years\cite{e}.
This suggests that the energy scale associated with baryon
number violating processes is greater than $10^{11}$ GeV
but remains below of the $10^{16}$ GeV.
Indeed, if the temperature of the
universe grows until values of temperature greater
than $\theta_B$ before the thermal equilibrium is restored,
fresh inflation could give violation of baryon number conservation.
During fresh inflation
the entropy density is $s \simeq {2\pi^2\over 45} g_{eff} \theta^3$.
If the decay of each $\phi$-boson on average a net baryon
number density $\epsilon $, then the net baryon
number density produced by the $\phi$-decay in unstable bosons is 
$n_B \simeq \epsilon n_{\phi}$, where $n_{\phi} \simeq
{\rho_r \over {\cal M}(\theta)}$, such
that
\begin{equation}
n_B \simeq \epsilon \frac{(n+2) \lambda^2\theta^2\phi^2}{
12 \sqrt{{\cal M}^2(0) + \frac{(n+2)}{12} \lambda^2 \theta^2}}.
\end{equation}

But the important relationship here is $n_B/s$, which takes the
form
\begin{equation}
\frac{n_B}{s} \simeq  \frac{45 \epsilon (n+2) \lambda^2\phi^2}{24\pi^2 g_{eff} \theta
\sqrt{{\cal M}^2(0)+\frac{(n+2)}{12} \lambda^2 \theta^2}}.
\end{equation}
If the temperature is given by\cite{22}
\begin{equation}
\theta(t)\simeq \frac{192 \pi}{g^4_{eff}} {\cal M}^2(0) \  t,
\end{equation}
the time for which the condition to baryogenesis take place
$n_B/s \simeq 10^{-10}$\cite{7} (I call it $t_B$), will be
\begin{equation}
t_B \simeq \frac{22}{\epsilon},
\end{equation}
where we have taken ${\cal M}^2(0) = 10^{-12} \  {\rm M^2_p}$ and
$g_{eff} = 10^{2}$.
Notice that $t_B$ is smaller than the time when the thermal equilibrium
is restored $t_E \simeq 10^{12} \  {\rm G^{1/2}}$, for
\begin{equation}
\epsilon < 10^{-11},
\end{equation}
which agree with the expected values.
Finally, the minimum radiation temperature needed to baryogenesis
take place (of the order of $10^{-8} \  {\rm G^{-1/2}}$) is obtained
for $t_B \simeq 10^{9} \  {\rm G^{1/2}}$, which is smaller than
the equilibrium temperature $t_E$. This implies the possibility
of violation of baryon number conservation during the out-of-equilibrium
period of fresh inflation.
Notice that when the thermal equilibrium is restored the temperature
of the universe is of the order of $\theta_E \simeq 10^{-6} \  {\rm G^{-1/2}}$,
which is in the permited range of temperatures
[$(10^{-8} - 10^{-3}) \ {\rm G^{-1/2}}$] for baryogenesis can take place.
In other words, with the choice of parameters here worked
(${\cal M}^2(0)=10^{-12} \  {\rm G^{-1}}$ and $g_{eff} = 10^{2}$),
the fresh inflationary
model predicts the possiblity of violation for baryon number conservation
in the range of times $(10^{9} - 10^{12}) \  {\rm G^{1/2}}$.

To summarize, the conditions needed to give rise to a baryon
asymmetry have long been recognized. They are (i) violation
of baryon number conservation (ii) violation of $CP$ invariance
and (iii) temporary loss of thermal equilibrium. 
Inflation requires violation of baryon number conservation, which
suggests that the proton is unstable. This provides us with
a bleak picture of a future universe devoid of matter an
ever decreasing photon density.
In this paper I have showed that baryogenesis can take place during fresh
inflation before the thermal equilibrium is restored.
This is a very attractive prediction of fresh inflation 
which shows
an important difference with respect to another models of inflation
where baryogenesis
is produced at the end of the inflationary phase\cite{ul,ul1}.
In the framework of fresh inflation, other interesting variants such
preheating\cite{KLS} or the Afflek-Dine mechanism\cite{AD} cannot occur due
to there are no oscillation of the inflaton field at the end
of fresh inflation.
Finally, baryogenesis appears to be very difficult in low-energy unification
scenarios and in supersymmetric unified models with dimension-$5$ operators
that violate $B$ conservation. However, while it is difficult to generate
a baryon asymmetry at low temperature, it is not impossible. A scenario
based on $SU(2)_L \otimes SU(2)_R \otimes U(1)_{B-L}$ where the
$X$-bosons are the right-handed neutrino and $M=10^4 \  {\rm GeV}$ has
been discussed in\cite{MM}.

\vskip .1cm
\centerline{{\bf ACKNOWLEDGMENTS}}
\vskip .1cm
\noindent
I would like to acknowledge CONACYT (M\'exico) and CIC of Universidad
Michoacana for financial support in the form of a research grant.\\

\end{document}